\documentclass[prd,aps,floatfix,nofootinbib,preprint ,tightenlines ]{revtex4}
\usepackage{dcolumn}
\usepackage{amsmath}
\usepackage{latexsym}
\usepackage{graphicx}
\usepackage{bm}

\long \def \blockcomment #1\endcomment{}

\newcommand{\bee}{\begin{equation}}
\newcommand{\ee}{\end{equation}}
\newcommand{\beea}{\begin{eqnarray}}
\newcommand{\eea}{\end{eqnarray}}

\begin{document}
\title{%
Remarks on lattice gauge theories with infrared-attractive fixed points}
\author{Thomas DeGrand, Anna Hasenfratz}%
\affiliation{Department of Physics,
University of Colorado, Boulder, CO 80309, USA}

\begin{abstract}
Theories of interacting gauge fields and fermions can possess 
a running gauge coupling with an infrared attractive fixed point (IRFP).
 We present a minimal description of the physics of these systems and
comment on some simple expectations for results  from lattice simulations
done within the basin of attraction of the IRFP in  these theories.
\end{abstract}

\maketitle

\section{Introduction}

Interest in non-Abelian lattice gauge theories which differ
 from QCD in having more quark flavors or quarks in larger representations 
has been growing in recent years.
These studies aim to address physics that may appear at the energy
of the Large Hadron Collider or beyond, described  by theoretical constructs such as
 technicolor \cite{Hill:2002ap}
 or ``unparticles'' \cite{Georgi:2007si}.

Most of the interesting theories are asymptotically free and an infinite cut-off
limit can be defined at the ultraviolet Gaussian fixed point (FP) at 
bare couplings $g^2=0$, $m=0$. Starting with bare parameters near the FP both the 
running gauge coupling and the mass
increase as the energy scale is lowered. As long as the  coupling $g^2$ 
 remains  small the perturbative renormalization 
group $\beta$ function describes its running. There are two possibilities
at stronger coupling\cite{Caswell:1974gg,Banks:1981nn}.
 The first is when the Gaussian fixed point is the only 
fixed point with a diverging correlation length. A familiar example is  QCD with two light flavors. 
The $\beta$ function stays negative even at strong coupling, chiral symmetry
is spontaneously broken and the model
is confining.   
The second possibility is that the $\beta$ function develops another
zero, corresponding to an infrared attractive  fixed point (IRFP). In this case
 the infrared physics of the massless theory
 will possess a conformal
 symmetry that precludes confinement and the spontaneous breaking of chiral symmetry.
These  are the "unparticle"  theories.
Just before the conformal window opens up there is a possibility that  the
$\beta$ function, while staying negative, becomes small, thus the running of the 
coupling slows down to "walking".  Such theories are leading candidates
for technicolor models.

Most of the interesting cases in the literature have nonperturbative dynamics.
The theoretical approaches used to analyze their behavior are not under complete control.
Accordingly, several groups have begun to use lattice methods to investigate these models.
In at least three cases (an $SU(3)$ gauge group with 12 flavors of fundamental
representation fermions\cite{Appelquist:2007hu,Appelquist:2009ty}, $SU(3)$ with two flavors of
 symmetric-representation fermions\cite{Shamir:2008pb},
and $SU(2)$ with two flavors of adjoint representation fermions\cite{Hietanen:2009az})
authors have reported evidence for an IRFP theory.

This short note is written to present the simplest theoretical description of IRFP
theories. While we believe that a considerable fraction of the lattice community is
familiar with  the physics we will describe, we have participated in enough conversations
 and read enough published papers in which the description we will give was 
incompletely presented,
that we feel a review might be useful. A large part of the non-lattice literature
about IRFP theories is devoted to describing the transition from an IRFP
system to a confining one, as the number of flavors of fermions is reduced
\cite{Cohen:1988sq,Appelquist:1996dq,Appelquist:1998rb,gardi,Kaplan:2009kr}.
Our ``theoretical minimum'' might be useful as a benchmark against which
simulation results and these predictions can be compared.

After some introductory remarks, we would like to make two points:
\begin{itemize}
\item When a theory is in the basin of attraction of an IRFP,
its  gauge coupling is irrelevant. The only relevant coupling, which controls the leading scaling 
behavior of the theory, is the fermion mass.
\item The value of the IRFP gauge coupling, $g^*$,  is scheme-dependent. Spectral observables
cannot depend on its value, and can only be sensitive to the extreme values of the bare gauge coupling
which mark the basin of attraction of the IRFP region. The renormalization group flow is always towards
$g^*$, but its absolute direction depends on the scheme-dependent location of $g^*$.
\end{itemize}
We now elaborate on these points and their consequences.

\section{Mapping the phase structure with the Renormalization Group}

The renormalization group (RG) transformation is a frequently used
tool in studying the infinite cut-off limit of quantum field theories.
There are two different approaches. The Callan-Symanzik equation describes
the problem from the perturbative point of  view and is most often
used in connection with continuum regularization schemes. It describes
the change of the parameters of the theory as the function of the
cut-off, or equivalently, the renormalization subtraction mass $\mu$.
The calculation is perturbative and only parameters already present
in the Lagrangian are considered. For example the $\beta$ function
\bee
\beta(g^{2})=\frac{dg^{2}}{d\log(\mu^{2})}
\ee
 describes the running of the coupling $g^{2}$, and there are similar
expressions for the mass and any other couplings present. Zeros of
the $\beta$ function correspond to either ultraviolet or infrared
fixed points (UVFP and IRFP), depending on the slope of $\beta(g^{2})$ 
(negative or positive, respectively).

The other option is the inherently non-perturbative  Wilson RG approach. 
There one considers
the evolution of all the possible couplings of the system under an
RG transformation that preserves the internal symmetries of the system
but integrates out the cut-off level UV modes. The
fixed points of the transformation are characterized by the number
of relevant couplings or operators, i.e. couplings that flow away
from the FP. Continuum (or infinite cut-off) limits are defined when
the relevant couplings are tuned towards the FP. Irrelevant couplings flow in the IR
to values which are independent of their UV values.
 The number of relevant
operators and their speed along the RG flow lines are universal properties,
related to the critical properties of the underlying continuum limit.
However the location of the FP is not physical, in fact different
RG transformations have different fixed points.

The relation between the Callan-Symanzik and Wilson RG approaches are straightforward:
the Callan-Symanzik $\beta$ function describes the RG flow of one or a few couplings,
usually implicitly in perturbation theory. The zeroes of the $\beta$ function correspond
 to the fixed points of the Wilson RG.
While the Callan-Symanzik equations can provide the first hint about
the phase structure,
unless the predictions are well controlled by perturbation
theory, a non-perturbative approach is needed to verify the existence
and study the properties of the relevant fixed points. Lattice calculations
provide an excellent tool for that. We believe that the Wilson RG
approach is better suited to interpret lattice results and in the
following we will use mainly that language.

\section{The phase diagram of gauge theories with many fermions}

Consider a theory with $SU(N_{c})$ gauge group, coupled to $N_{f}$
flavors of fermions in representation $R$. In our mind is a lattice
simulation and that is the phase diagram we will describe. The Gaussian
fixed point at $g^{2}=0$, $m=0$ is well understood perturbatively.
The mass is a relevant operator and it will presumably will remain
so even at strong gauge coupling. $m=0$ is a critical surface and
we want to investigate the running of the gauge coupling along it.
In the perturbative region this is described by the $\beta$ function
which is universal up to two loop level: 
\bee
\beta(g^{2})=\frac{dg^{2}}{d\log(\mu^{2})}=
\frac{b_{1}}{16\pi^{2}}g^{4}+\frac{b_{2}}{(16\pi^{2})^{2}}g^{6}+\dots\label{eq:betaQCD}
\ee
 \beea
b_{1} & = & -\frac{11}{3}N_{c}+\frac{4}{3}N_{f}T(R)\nonumber \\
b_{2} & = & -\frac{34}{3}N_{c}^{2}+N_{f}T(R)\big(\frac{20}{3}N_{c}+4C_{2}(R)\big)\,. \nonumber \\
\eea
If the fermion number $N_{f}\le (11/4)N_c/T(R)$ (16.5 for fundamental representation
fermions in QCD), $b_{1}<0$ and both the gauge
coupling and the mass are relevant operators at the Gaussian FP. The theory is asymptotically
free, the coupling $g^{2}$ increases as the lattice spacing (or inverse
cut-off) increases.
For small number of fermions the perturbative $\beta$ function remains
negative even at strong coupling. Lattice simulations with $N_{f}=2$, 2+1, and
to some extent up to $N_f\le 8$  verify this: QCD with 2-8 light fermions is confining
and chirally broken everywhere \cite{severaln8}. With increasing  $N_{f}$  the two-loop
term in Eq.~\ref{eq:betaQCD} changes sign, suggesting a possible
zero in $\beta(g)$. When the quarks are in higher representations
of the gauge group, this happens more easily. The zero of $\beta(g)$
 corresponds to an IRFP in the gauge coupling. That is,
the gauge coupling is irrelevant, and the only relevant parameter is the
mass term. The continuum limit is defined in the basin of attraction of
 this IRFP by tuning the mass to zero.

\begin{figure}
\begin{center}
\includegraphics[width=0.5\textwidth,clip]{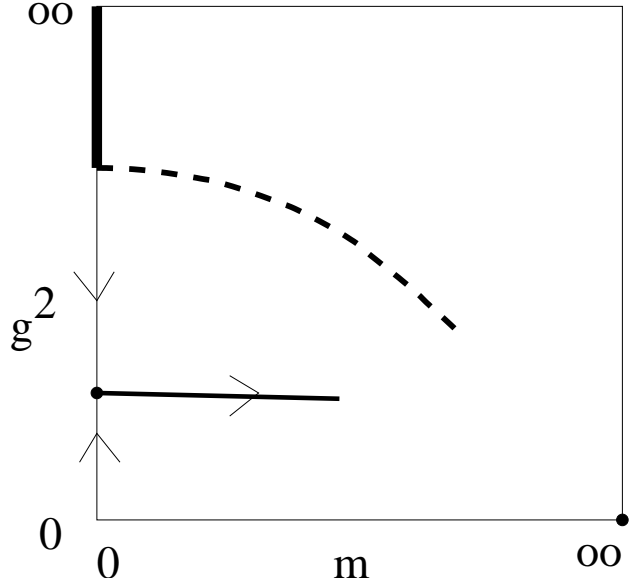}
\end{center}
\caption{ A conjectured  phase diagram for a lattice theory with an IRFP. The thick line
at $m=0$
is the confining, strong-coupling region of the massless theory. The dotted line separates
a confining phase (which includes the entire $g^2$ line at large fermion mass) from
the basin of attraction of the IRFP. Arrows show some IR flow lines
and the renormalized trajectory extending from the critical point.}
\label{fig:irfp2}
\end{figure}

The existence of the  perturbatively predicted IRFP  and the properties of the theory 
 at even stronger coupling have
 to be studied with non-perturbative methods, like lattice simulations. The lattice provides an 
UV regularization scheme which is equivalent to continuum schemes at weak coupling. 
At strong gauge coupling lattice artifacts can completely change the system. 
It is generally believed that in the strong coupling limit
lattice models are always confining and chirally broken, independent of $N_f$  
\cite{Casher:1979vw,Greensite:1980hy,Kawamoto:1981hw}, This was observed in early $N_f=16$
simulations \cite{Heller:1997vh,Damgaard:1997ut},
 though numerical results in 
Ref.~\cite{Iwasaki:2003de} contradict this. In any case, the details in the strong coupling 
region are not universal and, for now, not very important either. Fig.~\ref{fig:irfp2} shows a
conjectured  phase diagram in the $g^{2}$
vs $m$ plane when an IRFP exists. The thick line at large $g^2$ shows the (possible) 
confining and chirally broken phase, but around the IRFP the theory
is chirally symmetric and deconfined. The dashed line indicates the phase boundary
between the two phases. It is likely only a crossover at finite mass. 
We note that in a recent paper it was suggested that 
in these many-fermion theories with an IRFP there is always another FP 
(zero of the $\beta$ function) 
at a stronger coupling \cite{Kaplan:2009kr}. If this conjecture is valid, the phase
 diagram has another UVFP at $m=0$.
This could be the boundary between the conformal and strong coupling phases in
 Fig.~\ref{fig:irfp2}, or an entirely new critical point. 

It is important to note that while at a UVFP
an observable phase transition occurs, there is nothing observable
at the IRFP. The location of an IRFP  on  the critical
surface is not physical. It depends on the specific renormalization group
transformation, and therefore no physical observable can identify it.
In particular, this means that  studies of candidate theories which only measure
spectroscopic observables cannot directly detect coupling constant flow
through the dependence of observables on irrelevant bare couplings. They can only detect the
perimeter of the basin of attraction of the FP.

 \section{Analysis of the phase diagram}

To begin, note first that the gauge coupling plays a very different role
around the UVFP  and in the vicinity of the  IRFP or trivial phases: at the UVFP
$g$ is relevant, and to drive the correlation length(s) to infinity in units of the 
cutoff scale involves a fine tuning of $g$ toward zero.
(To be precise, the gauge coupling is marginally relevant with respect to
the Gaussian fixed point, because $\beta(g^2) \sim b_1 g^4$.)  In the other phases, IR physics
is independent of the value of $g$ at the UV scale, because the IR flow of the gauge coupling
is into its fixed point value:  the gauge coupling is irrelevant.
This is a simple example illustrating the general feature, that relevance or irrelevance
of an operator is measured with respect to a fixed point. The presence of the IRFP means that
 the linearized behavior of the beta function near the fixed point is
\bee
\frac{dg^2}{ds} = y_g(g^2-g^{2*}) ,
\label{eq:lbf}
\ee
where $s$ is the (IR) scale change, $\mu \to e^{-s}\mu$, and $y_g<0$ is the scaling exponent of
the gauge coupling. The running coupling then is
 $ g^2(s) = g^{2*} + (g_0^2-g^{2*}) e^{y_g s }$, i.e. $g(s)\to g^* $ as $s\to\infty$.
We discuss the value of $y_g$ below.

Consider next the general situation, a theory in $D$ Euclidean dimensions defined by a
 set of scaling operators $\{u\}$.
Under a real-space blocking by a factor $b$, the operators run multiplicatively to/from their FP
values, $u_i' -u_{i0} =
b^{y_i} (u_i - u_{i0})$, with the usual assignment
of relevancy, irrelevancy, or marginality, depending on whether $y_i$ is positive, negative,
or zero.  Let us assume that the mass is  the only relevant coupling
and its FP value is zero; its exponent we will label as $y_m$.
Only the leading exponent governs the
 correlation length $\xi$, which  diverges with the usual algebraic behavior
\bee
\xi \sim m^{-\frac{1}{y_m}}.
\label{eq:corrlen}
\ee
A standard textbook analysis tells us that the
the singular part of the free energy per site
(it is the part of the free energy containing all the non-analyticity at the critical point)
scales as
\bee
f_s(m,u_i) = m^{D/y_m} f_s( m_0,u_{i0} +
(u_i-u_{i0})\left(\frac{m}{m_0}\right)^{|y_i|/y_m}),
\label{eq:five}
\ee
which can be Taylor expanded as
\bee
f_s(m) = m^{D/y_m}(A_1 + A_2  m^{|y_i|/y_m}),
\label{eq:fs}
\ee
where $A_1$ and $A_2$ are non-universal constants. All other observables have a similar expansion.

Precisely at $m=0$ the entire basin of attraction of the FP is critical --
at long distance all correlation functions decay algebraically
\bee
\langle \phi_i(r) \phi_i(0)\rangle = \sum_j \frac{E_j}{r^{2(D-y_j)}}.
\label{eq:corrfn}
\ee
This behavior is only achieved by
setting the quark mass to zero. The nonanalytic dependence of the free energy 
on the fermion mass
is responsible for the algebraic behavior of correlation functions in the massless theory.
 Eq.~\ref{eq:corrfn}  is only true asymptotically. At distances which are comparable
to the UV cutoff (lattice spacing $a$) there are additional contributions.
They arise from the non-singular part of the free energy, from physics at the intermediate
scales which has been integrated out in the construction of Eq.~\ref{eq:five}.
These give extra contributions going as $\exp(-r/a)$.

On the  $m=0$ critical surface all couplings are irrelevant; they all flow into the IRFP.
The interesting physical quantity near the FP is the leading relevant exponent. This is not
accessible from a lattice-based RG study at $m=0$ (for example, a conventional Schr\"odinger functional
study which computes $\beta(g)$).
These studies only give $y_g$ because by construction they are
 only  sensitive to flow into the FP. They tell us that there
is an IRFP, no more.

Unitarity bounds for $D=4$ conformal field theories \cite{Mack:1976pa} constrain the scaling dimension of
the condensate $\langle \bar \psi \psi \rangle \sim \mu^\gamma$ to lie in the range $3>\gamma>1$.
Since $m \bar \psi \psi$ is scale invariant and since all dimensionful quantities scale
with the correlation length as in Eq.~\ref{eq:corrlen}, $\gamma=4-y_m$.
(There are other definitions for the scaling dimension in the literature but this one is consistent
with the bound of Ref.~\cite{Mack:1976pa}.)
We do not want to discuss specific
techniques for determining it in a simulation, 
and  defer a discussion of possible values for $y_m$ to later work \cite{Hasenfratz:2009ea}.
In the meantime, however, there
 is one case for which it is easy to expose the operator scaling hierarchy 
\cite{Banks:1981nn,Grinstein:2008qk}. This is a
 lattice theory of $N_c$ colors and $N_f$
flavors in some representation, and that the limit $N_c \rightarrow\infty$ and $N_f/N_c$ fixed,
 so that the
Banks-Zaks FP occurs at a small value of $g^*\sim \epsilon$. Engineering dimensions give
basically everything.
For this theory, the condensate has scaling exponent $\gamma=3 - O(\epsilon)$.
The gauge coupling contributes a scaling exponent $y_g\sim \epsilon$ (which can be computed
in perturbation theory), and lattice-based artifacts
contribute negative $y_i$'s.

This is the upper end of the conformal window. What happens at the lower end of the
conformal window, where the system converts from being an IRFP theory to a confining, chirally
broken theory, is a subject of long-standing interest. 
Observations of $y_m$ will likely be needed to elucidate this point.

\section{Consequences for simulations}
One feature of IRFP theories is concealed by the Callan-Symanzik formalism: the location of
a fixed point is not universal. Different choices of renormalization group transformations
 will result in different
values of $g^*$. This has several consequences for simulations.

First of all,  observables taken from correlation functions cannot depend on
the actual value of the FP coupling determined in some scheme: the correlation function
does not know about the scheme. The important part of expressions
 like Eq.~\ref{eq:fs} is the statement
that the size of scale violations in observables is
$\left(\frac{m}{m_0}\right)^{|y_i|/y_m}$.

Real QCD (for example, an $SU(3)$ gauge theory with a small number of flavors of
 fundamental representation
 fermions)
is more complicated than an IRFP theory. Now there are two relevant couplings which flow away from
the $(g^2, m_q)=(0,0)$ fixed point. Lattice observables (like masses) depend strongly
on both of these couplings. Indeed, making lattice predictions typically involves first
finding a map
of $m_q(g^2)$ along which some mass ratio is fixed, then taking the continuum limit by moving
along this line toward the UVFP.
If one tried to do this for an IRFP theory, one would be trying to express (or tune) the relevant coupling
in terms of an irrelevant coupling.

Next, because different choices for renormalization schemes result in different
values for $g^*$, it is easy to imagine a situation where the running coupling in one scheme 
is increasing under flow to the IR, while the running coupling in another scheme is decreasing.
In fact, this is another diagnostic for the presence of an IRFP: one can vary the
renormalization
group transformation and see
if the direction of coupling constant flow changes.

Much continuum beyond-Standard Model phenomenology uses a scale dependent coupling constant
defined in some particular scheme, such as $\overline{MS}$. A lattice calculation of such a coupling
involves both a lattice simulation and an additional scheme matching calculation.
Critital exponents such as $y_m$ are scheme-independent.

Simulations are carried out at particular values of the bare couplings.
Eq.~\ref{eq:fs} tells us that the influence of the gauge coupling on observables dies away
in the zero mass limit.
In a sense, varying the gauge coupling while remaining near the basin of attraction of the IRFP
is like doing lattice simulations for ordinary QCD with several choices of the lattice action --
in the scaling limit, the answer is not supposed to depend on the choice of action.
Away from infinite correlation length, different lattice actions will give different predictions
 for observables. But these differences are just scale violations. In this case, the choice of
bare gauge coupling is as much a choice of the action as a particular discretization would be.

The correlation length $\xi$ 
 only varies with the bare mass according to Eq.~\ref{eq:corrlen} when the system size $L$ is much larger
than $\xi$.  This is probably the the most serious practical constraint on the choice of bare parameters.

The size of the exponent $y_g$ also acts to minimize the dependence of simulation results on $g^2$:
It is likely the value of $y_g$ is very small.  The exponent
 can be seen in simulations. The authors of Ref.~\cite{Appelquist:2009ty}
report $y_g \sim -0.15$. The authors of Ref.~\cite{Hietanen:2009az} do not quote a number, but
they give a plot of the $\beta$ function from which a value of about -0.16 can be inferred.
The authors of Ref.~\cite{Shamir:2008pb} only show an integrated beta function, but parameterizing their
data with a simple model (a second order formula with the perturbative $b_1$ and a $b_2$ chosen
to give a zero) also gives a small $y_g \sim -0.06$. Gardi and Grunberg\cite{gardi} present 
perturbative calculations for both QCD-like and supersymmetric theories. Only at the bottom of
the conformal window, and only for non-supersymmetric theories, is $-y_g$ greater than about 0.5.
If that is the case, then
in formulas like Eq.~\ref{eq:fs} the correction term
is nearly a constant. 
This has two consequences:
\begin{itemize}
\item In a lattice simulation of a particular system
which is a candidate for being an IRFP theory,
the observation that long distance observables become independent of $g$ at small $m$ is a positive
indication of IRFP behavior
\item In the basin of attraction of an IRFP, one does not need to fine tune the gauge coupling;
one can compute wherever it is convenient.
\end{itemize}

What evidence  is there for the behavior we have described, from published results for candidate
IRFP theories? 
Two groups have studied the $SU(2)$ gauge group with two flavors of adjoint fermions.
This model has a confining strong coupling phase and a weak coupling phase which the
authors of Ref.~\cite{Hietanen:2009az}  have argued is conformal.
The authors of Ref.~\cite{Catterall:2008qk} measure the pseudoscalar
and vector masses in finite volume at zero quark mass. Finite volume limits
the correlation length to be proportional to the system size. They observe (in their Fig. 5) that
 these masses are independent of the bare gauge coupling in the weak coupling phase.
 They also note the constancy of Creutz ratios with respect to gauge coupling
 (ratios of
Wilson loops) in this phase; see their Fig. 7. The authors of Ref.~\cite{Hietanen:2008mr}
observe that their pseudoscalar and vector masses are independent of gauge coupling
in the weak coupling phase; compare their figures 5-8.

\begin{figure}
\begin{center}
\includegraphics[width=0.7\textwidth,clip]{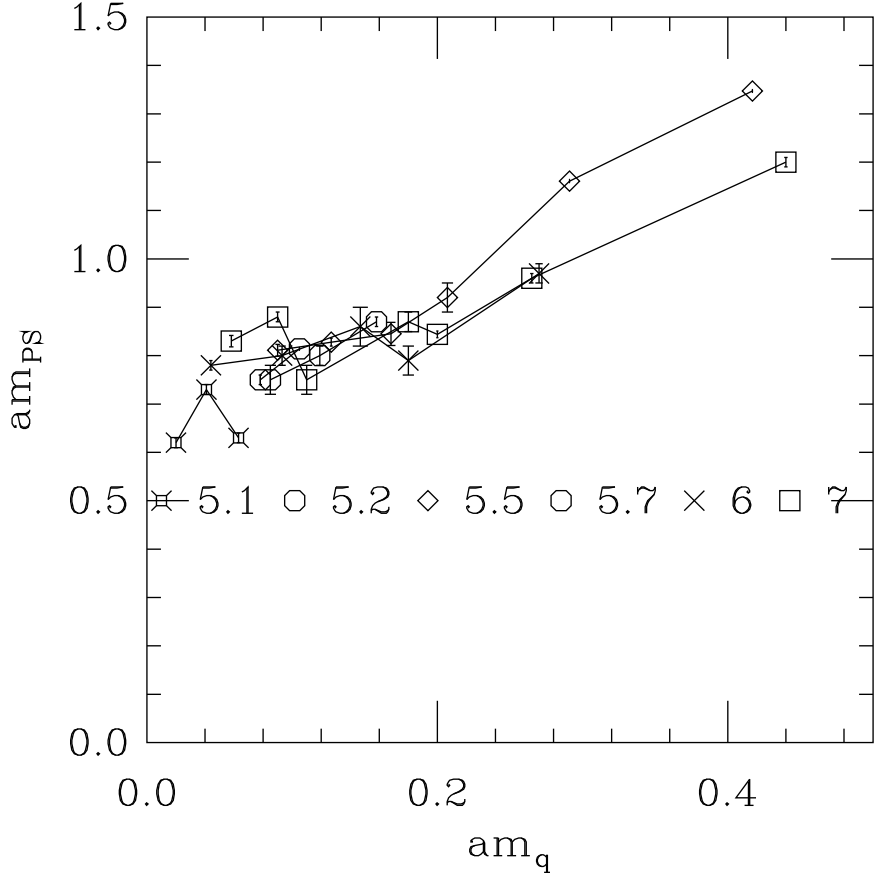}
\end{center}
\caption{Replottted data from Ref.~\protect{\cite{DeGrand:2008kx}} for the pseudoscalar
screening mass on $(12\times 8^2)\times 8$ lattices, in the weak coupling phase
of $SU(3)$ gauge theory with two flavors of sextet representation quarks.
Lines join the equal gauge coupling data, with different plotting symbols for the different $\beta$ values.
}
\label{fig:mpi1288}
\end{figure}

One of us collaborated with Svetitsky and Shamir in studies of $SU(3)$ gauge theory
with two flavors of sextet fermions \cite{DeGrand:2008kx}.
 This system also has a weak coupling phase which is a 
candidate for an IRFP phase. It is easier to replot data published there: Fig. \ref{fig:mpi1288}
shows the pseudoscalar mass as a function of the quark mass, from simulations performed at
 many values of the gauge coupling ($\beta=6/g^2$) in the weak coupling phase.
The finite simulation volume prevents the vanishing of the pseudoscalar mass at zero quark mass.
Results are similar to what is seen in the $SU(2)$, adjoint case, i.e., weak dependence
of the meson mass on the gauge coupling.

\section{Conclusions}
We have presented a minimal scenario for physics within the basin of attraction of an IRFP.
There is one relevant coupling, the fermion mass, which governs the correlation length
through Eq.~\ref{eq:corrlen}.
All other couplings are irrelevant. This scenario gives a context for interpreting simulation
 results
and has observable consequences. Probably, convincing evidence for an IRFP requires
RG studies. However, the simple behavior we have described 
provides  additional markers for the description of an IRFP theory with a
 single relevant operator.

\begin{acknowledgments}
We thank P.~Damgaard, U.~M.~Heller, D.~M.~Kaplan,  T.~G.~Kovacs,
L.~Del Debbio,
 B.~Svetitsky and Y.~Shamir for discussions.
This work was supported in part by the US Department of Energy.

\end{acknowledgments}

\end{document}